\documentclass[preprint,12pt]{elsarticle}
\usepackage{amssymb}

\begin{document}
\begin{frontmatter}

\title{A single trapped ion in a finite range trap}

\author[]{M. Bagheri Harouni \corref{cor1}}
\ead{m-bagheri@phys.ui.ac.ir}
\author[]{M. Davoudi Darareh}
\ead{m.davoudi@sci.ui.ac.ir}
\address{Department of Physics,
Faculty of Science, University of Isfahan, Hezar Jerib, Isfahan,
81746-73441, Iran} \cortext[cor1]{Corresponding author. Tel.:+98
311 7932435 ; fax: +98 311 7932409
 E-mail address: m-bagheri@phys.ui.ac.ir (M. Bagheri Harouni)}

\begin{abstract}
This paper presents a method to describe dynamics of an ion
confined in a realistic finite range trap. We model this realistic
potential with a solvable one and we obtain dynamical variables
(raising and lowering operators) of this potential. We consider
coherent interaction of this confined ion in a finite range trap
and we show that its center-of-mass motion steady state is a
special kind of nonlinear coherent states. Physical properties of
this state and their dependence on the finite range of potential
are studied.
\end{abstract}


\end{frontmatter}

\section{Introduction}
\indent Single trapped ions represent elementary quantum systems
that are approximately isolated from the environment
\cite{leibfried}. In these systems both internal electronic states
and external center-of-mass motional states (external states) can
be coupled to and manipulated by light fields. This makes the
trapped ion systems suited for quantum optical and quantum
dynamical studies under well-controlled conditions. Motivated by
the strong analogy between cavity quantum electrodynamics and the
trapped ion system, various theoretical and experimental proposals
have been made on how to create nonclassical and arbitrary states
of motion of trapped ions. Preparation of the number states
\cite{roos}, coherent, quadrature squeezed number states and
superposition of the number states were considered in this system
experimentally \cite{meekhof} and theoretically \cite{heinzen}.
Experimental preparation of the Schr\"{o}dinger-cat state was
considered \cite{monroe1}. Theoretical schemes for generation of
arbitrary center-of-mass motional states of a trapped ion is
described in \cite{gardiner}. Moreover, the possibility of
generation of even and odd coherent states of the center-of-mass
motion of a trapped ion is considered \cite{filho}. A new scheme
for preparation of nonclassical motional states of trapped ions is
investigated in \cite{wang}. Recently, Preparation of Dicke states
in an ion chain is considered theoretically and experimentally
\cite{hume}. In addition to the above-mentioned attempts,
preparation of different family of nonlinear coherent states
\cite{manko}, is also studied theoretically \cite{vogel1,mahd}.\\
\indent On the other hand, the trapped ion system has found some
applications in quantum information and quantum computation
\cite{doerk}. For quantum information processing by trapped ion,
preparation of some special states is important. Among these
states, entangled states have found crucial importance.
Preparation of the entangled states of trapped ion is considered
recently \cite{blatt}. For quantum computation applications,
preparation of the two-dimensional cluster-state is considered
\cite{wunderlich}. Because of the some similarities between the
trapped ion system and the Jaynes-Cummings model \cite{jaynes},
the trapped ion system is used to realize different
generalizations of the Jaynes-Cummings model which have found
some applications in quantum information \cite{militello}.\\
\indent In all of the above-mentioned efforts on the trapped ion
system, it is assumed that the ion is confined in a harmonic
oscillator-shaped potential while the dimension of this potential
extended to infinity. Hence, the range of the confining trap is
infinity. However, in the realistic experimental setup, the
dimension of trap is finite and the realistic trapping potential
is not a harmonic oscillator potential but the truncated and
modified one within the extension of the trap. In this paper, we
assume that the confining potential for ion has finite range.  We
will model this confining potential with a solvable one. By using
the concept of the $f$-deformed oscillator \cite{manko}, we try to
consider the trapped ion in confining potential with finite range
as an $f$-deformed oscillator and in this context we obtain
raising and lowering operators (dynamical variables) of this
potential. The finite range effects of this model can be used in
traps of the order of nano-scale, called nano Paul traps, that are
attracted a great deal of attention recently \cite{nanotrap}. It
is worth to note that the confining model potential which is
considered here is used for other confined physical systems, such
as the Bose-Einstein condensate \cite{wang1}, and carriers in a
quantum well \cite{harrison}. The $f$-deformed oscillator approach
where we have considered here, has been used before for some other
confined systems \cite{malek1}.\\ \indent This paper presents a
method to describe dynamics of an ion confined in a finite range
trap. We will show that stationary state of the center-of-mass
motion of the trapped ion is a special kind of nonlinear coherent
states where its properties depend on the range of the confining
potential. The outline of the paper is as follows. Section
\ref{sec1} deals with scheme for model potential and in this
section we will obtain dynamical variables of this potential in
the context of the $f$-deformed oscillator. In Sec. \ref{sec2} we
propose coherent interaction of an ion confined in a finite range
potential and we consider its dynamics in the steady state. In
this section we will obtain an eigenvalue equation for the state
of the center-of-mass motion of the ion. In Sec. \ref{sec3} we
summarize definition of the nonlinear coherent states and we will
show that the steady state of the ion motion can be considered as
a nonlinear coherent state. Physical properties of this system are
investigated in this section. Section \ref{sec4} is devoted to the
conclusion.
%-------------------------------------------------------------------------------------
\section{Algebraic approach for a particle in a finite range potential}\label{sec1}
\indent To consider an ion in a finite range trap, we try to model
the potential energy function of the realistic trap by an
analytically solvable potential. For comparing new results with
previous ones we are looking for a potential which reduces to the
harmonic oscillator potential in a specific limit of its
parameters. A potential which has this property is the modified
P\"{o}schl-Teller (MPT) potential ~\cite{MPT}. The MPT potential
has the following form
\begin{equation}\label{mpt potential}
V(x)=D\,tanh^2(\frac{x}{\delta}),
\end{equation}
where $D$ is the depth of the well, $\delta$ determines the range
of the potential and $x$ gives the relative distance from the
equilibrium position. The well depth, D,  can  be defined as
$D=\frac{1}{2}m\omega^2\delta^2$, with mass of the particle $m$
and angular frequency $\omega$ of the harmonic oscillator, so
that, in the limiting case $D\rightarrow \infty$(or $
\delta\rightarrow \infty)$, but keeping the product $m\omega^2$
finite, the MPT potential energy reduces to the harmonic potential
energy,  $\lim_{D\rightarrow \infty}
V(x)=\frac{1}{2}m\omega^2x^2$. Solving the Schr\"{o}dinger
equation, the energy eigenvalues for the MPT potential are
obtained as~\cite{landau}
\begin{equation}\label{mpt energy 1}
E_n=D-\frac{\hbar^2\omega^2}{4D}(s-n)^2,     \hspace{1cm} n=0, 1,
2, \cdots, [s]
\end{equation}
in which $s=(\sqrt{1+(\frac{4D}{\hbar\omega})^2}-1)/2$,  and $[s]$
stands for the closest integer to $s$ that is smaller than $s$.
The MPT oscillator quantum number $n$ can not be larger than the
maximum number of bound states $[s]$, because of the dissociation
condition $s-n\geq 0$. Detailed description about this energy
spectrum can be found in ~\cite{Davoudi}. By introducing a
dimensionless parameter
$N=\frac{4D}{\hbar\omega}=\frac{2m\omega\delta^2}{\hbar}$, the
bound energy spectrum in equation (\ref{mpt energy 1}) can be
rewritten as
\begin{equation}\label{mpt energy 2}
E_n=\hbar\omega[-\frac{n^2}{N}+(\sqrt{1+\frac{1}{N^2}}-\frac{1}{N})n+
\frac{1}{2}(\sqrt{1+\frac{1}{N^2}}-\frac{1}{N})].
\end{equation}
The relation (\ref{mpt energy 2}) shows a nonlinear dependence on
the quantum number $n$, so that, different energy levels are not
equally spaced. As is evident, $N$ is a dimensionless parameter
and from now we refer this parameter as the depth of the trap. It
is clear that, in the limiting case $D\rightarrow \infty$ (or
$N\rightarrow \infty$), the energy spectrum for the quantum
harmonic oscillator will be obtained, i.e.,
$E_n=\hbar\omega(n+\frac{1}{2})$. This means that for finite
values of $D$ (or finite values of $\delta$), we have a deformed
quantum oscillator, which its natural deformation from the quantum
harmonic oscillator can be amplified by decreasing $D$ or $N$.
Thus, the well depth of this potential that identifies its range,
is used to approximate the harmonic oscillator potential and it
can also be considered as a controllable physical deformation
parameter. It is interesting to note that the dimensionless
parameter $N$ can also be written as $N=\frac{\delta}{\Delta x}$.
Here $\Delta x=\sqrt{\frac{\hbar}{2m\omega}}$, is the ground state
wave function spread which for typical traps is of the order of
nanometer $(nm)$ \cite{leibfried}. $\delta$, that determines the
range of the potential would be of the same order of magnitude as
the ion-electrode distance in a Paul trap system. It results that
if trap size be of the order of $nm$, the finite range effects of
the trap would be important. Such kind of the Paul traps have
considered recently \cite{nanotrap}.\\
\indent It is shown that \cite{malek1}, each quantum system which
has an unequal spaced energy spectrum can be considered as an
$f$-deformed oscillator. Therefore, according to the energy
spectrum of the MPT potential, this system can be considered as an
$f$-deformed oscillator ~\cite{Davoudi}. On the other hand, the
$f$-deformed quantum oscillator \cite{manko}, as a nonlinear
oscillator with a specific kind of nonlinearity, is characterized
by the following deformed dynamical variables $\hat{A}$ and
$\hat{A}^\dag$
\begin{eqnarray}\label{fd}
\hat{A}&=&\hat{a}f(\hat{n})=f(\hat{n}+1)\hat{a},\nonumber\\
\hat{A}^\dag&=&f(\hat{n})\hat{a}^\dag=\hat{a}^\dag f(\hat{n}+1),
\hspace{1.5cm} \hat{n}= \hat{a}^\dag\hat{a},
\end{eqnarray}
\noindent where $\hat{a}$ and  $\hat{a}^\dag$ are usual boson
annihilation and creation operators $([\hat{a}, \hat{a}^\dag]=1)$,
respectively. The real deformation function $f(\hat{n})$ is a
nonlinear operator-valued function of the harmonic number operator
$\hat{n}$, which introduces some nonlinearities to the system.
From equation (\ref{fd}), it follows that the $f$-deformed
operators $\hat{A}$, $\hat{A}^\dag$ and $\hat{n}$ satisfy the
following closed algebra
\begin{eqnarray}\label{algebrafd}
&[\hat{A}, \hat{A}^\dag]=&(\hat{n}+1)f^2(\hat{n}+1)-\hat{n}f^2(\hat{n}),\nonumber\\
&[\hat{n}, \hat{A}]=&-\hat{A},\hspace{1.5cm} [\hat{n},
\hat{A}^{\dag}]=\hat{A}^{\dag}.
\end{eqnarray}
\noindent The above-mentioned  algebra, represents a deformed
Heisenberg-Weyl algebra whose nature depends on the nonlinear
deformation function $f(\hat{n})$. An $f$-deformed oscillator is a
nonlinear  system characterized by a Hamiltonian of the harmonic
oscillator form
\begin{equation}\label{hamiltf}
\hat{H}=\frac{\hbar\omega}{2}(\hat{A}^\dag\hat{A}+\hat{A}\hat{A}^\dag).
\end{equation}
Using equation (\ref{fd}) and the number state representation
$\hat{n}|n \rangle=n|n \rangle$, the eigenvalues of the
Hamiltonian (\ref{hamiltf}) can be written as
\begin{equation}\label{energyf}
E_n=\frac{\hbar\omega}{2}[(n+1)f^2(n+1)+nf^2(n)].
\end{equation}
\indent It is worth  noting that in the limiting  case
$f(n)\rightarrow 1$, the deformed algebra (\ref{algebrafd}) and
the deformed energy eigenvalues (\ref{energyf}) will reduce to the
conventional Heisenberg-Weyl algebra and the harmonic oscillator spectrum, respectively.\\
\indent Comparing the bound energy spectrum of the MPT oscillator,
equation (\ref{mpt energy 2}), and the energy spectrum of an
$f$-deformed oscillator, equation (\ref{energyf}), we obtain the
corresponding deformation function for the MPT oscillator as
\begin{equation}\label{ff}
f^2(\hat{n})=\sqrt{1+\frac{1}{N^2}}-\frac{\hat{n}}{N}.
\end{equation}
Furthermore, the ladder operators of the bound eigenstates of the
MPT Hamiltonian can be written in terms of the conventional
operators $\hat{a}$ and $\hat{a}^\dag$ as follows
\begin{equation}\label{a mpt}
\hat{A}=\hat{a}\sqrt{\sqrt{1+\frac{1}{N^2}}-\frac{\hat{n}}{N}},
\hspace{0.5cm}
\hat{A}^\dag=\sqrt{\sqrt{1+\frac{1}{N^2}}-\frac{\hat{n}}{N}}\hat{a}^\dag.
\end{equation}
\noindent These two operators satisfy the deformed Heisenberg-Weyl
commutation relation
\begin{equation}\label{algebra mpt}
[\hat{A},
\hat{A}^\dag]=\sqrt{1+\frac{1}{N^2}}-\frac{2\hat{n}+1}{N},
\end{equation}
As is clear, in the limiting case $f(n)\rightarrow
1\;(N\rightarrow\infty)$ this deformed commutation relation will
reduce to the conventional commutation relation,
$[\hat{a},\hat{a}^{\dag}]=1$.\\ \indent As a result, in this
section we conclude that the trapped ion in MPT potential can be
considered as an $f$-deformed oscillator with specific kind of the
$f$-deformed Heisenberg-Weyl algebra.\\ \indent In the following,
we will consider coherent interaction of a single trapped ion in a
finite range trap with light fields. Then, we will generate the
nonlinear coherent states of ionic vibrational motion in a finite
range trap and finally we will investigate some physical
properties of these states such as, their number distribution,
quadrature squeezing and their phase-space distribution.
%----------------------------------------------------------------------------------------
\section{Ion dynamics in a finite range trap}\label{sec2}
\indent As is usual in theoretical consideration of trapped ion
systems, the confining potential is assumed to be a spatial
varying high-frequency time-dependent field, the so-called Paul
trap, $V(\vec r,t)$. It is shown that, motion of a particle inside
a such high-frequency trap can be treated by averaging over the
fast motion (part of the particle displacement that its frequency
is the same as frequency of trap fields). In this approach a
confined particle in such a trap experiences a spatial static
effective potential \cite{landau1}. Usually this static potential
is assumed to be a three dimensional harmonic oscillator-like
potential so that in one direction ($x$-direction) can be written
as $V(x)=\frac{1}{2}m\omega^2x^2$ \cite{leibfried}. As is
conventional, ion is cooled to the ground-state of the trap and in
this situation due to smallness of the ratio of trap height to
other energy scales, such as energy distance between two adjacent
energy levels of the trap, the trap is assumed extend to infinity.
However, in the realistic experimental setup, the dimension of the
trap is finite and the realistic trapping potential is not a
harmonic oscillator potential extending to infinity but the
truncated and modified one within the extension of the trap. Thus,
the realistic confining potential becomes flat near the edge of
the trap and can be simulated by the tanh-shaped potential, so
that in one dimension ($x$-direction) can be written as
$V(x)=D\tanh^2(\frac{x}{\delta})$. In this paper, we try to
investigate some effects which originate from finite range
property of the trap.\\ \indent According to the previous section,
we model this trapped ion as an $f$-deformed quantum oscillator.
Therefore, the oscillator-like Hamiltonian of this system can be
written as
\begin{equation}\label{heh}
    \hat{H}_t=\frac{\hbar\omega}{2}(\hat A\hat A^\dag+\hat A^\dag\hat A),
\end{equation}
where we interpret the operator $\hat A\;(\hat A^\dag)$ as the
operator whose action causes the transition of the ion
center-of-mass motion to the lower (upper) energy state of the
trap. These operators are given in Eq. (\ref{a mpt}). In fact, the
Hamiltonian (\ref{heh}) is related to the external degrees of
freedom of the ion. According to the resonant condition, the ion
is assumed as a two-level system with the ground state $|g\rangle$
and the excited state $|e\rangle$. Then, internal degrees of
freedom of the ion can be expressed with electronic flip operators
$\hat S_z=|e\rangle\langle e|-|g\rangle\langle g|$, $\hat
S^+=|e\rangle\langle g|$ and $\hat S^-=|g\rangle\langle e|$ which
satisfy the usual $su(2)$ algebra. On the other hand, with the
help of the suitable laser fields, the internal levels of the
trapped ion can be coherently coupled to each other and to the
external motional degrees of freedom of the ion. Therefore, the
total Hamiltonian of the system may be given as
\begin{equation}\label{}
    \hat H=\hat H_0+\hat H_{int}(t),
\end{equation}
where $\hat H_0=\hat H_t+\hbar\omega_i\hat S_z$, with $H_t$ given
in Eq. (\ref{heh}), describes the free motion of the internal and
external degrees of freedom of the ion. Here, $\hbar\omega_i$
refers to the energy difference of internal states of the ion,
$\hbar\omega_i=E_e-E_g$. The interaction of the ion with the laser
fields is described by $\hat H_{int}(t)$ and is written as
\begin{equation}\label{hint}
    \hat H_{int}(t)=g\left[E_0e^{-i(k_0\hat x-\omega_it)}+E_1e^{-i[k_1\hat
    x-(\omega_i-\omega_n)t]}\right]\hat S^+ +\;H.c.\;,
\end{equation}
in which $g$ is coupling constant, $k_0$ and $k_1$ are the wave
numbers of the driving laser fields and $\omega_n$ refers to the
energy of the lower vibrational side-band with respect to the
electronic transition of the ion. $\omega_n$ is the frequency of
the ion transition between energy levels of the finite range trap.
Because energy spectrum of the trap depends on the energy level
numbers and we consider a transition between specific side-band
levels, hence, we show the transition frequency with definite
dependence to $n$. In the above Hamiltonian, $\hat H_{int}(t)$,
$\hat x$ is the operator of the center-of-mass position and may be
defined as \cite{malek1}
\begin{equation}\label{}
    \hat x=\frac{\eta}{k_l}(\hat A+\hat A^\dag),
\end{equation}
where $\eta$ being the Lamb-Dicke parameter and $k_l$ is
associated wave number to the characteristic length of the trap
and assume to be $k_l\simeq k_0\simeq k_1$. The interaction
Hamiltonian (\ref{hint}) can be written as
\begin{equation}\label{}
    \hat
    H_{int}(t)=\hbar e^{i\omega_it}\left[\Omega_0+\Omega_1e^{-i\omega_nt}\right]e^{i\eta(\hat A+\hat
    A^\dag)}\hat S^++H.c.\;,
\end{equation}
$\Omega_0=\frac{gE_0}{\hbar}$ and $\Omega_1=\frac{gE_1}{\hbar}$
are the Rabi frequencies of the laser fields tuned to the
electronic transition and the lower sideband, respectively. The
interaction Hamiltonian in the interaction picture with respect to
the $\hat H_0$ can be written as
\begin{equation}\label{hint1}
    \hat H_I=\hbar\Omega_1\hat
    S^+\left[\frac{\Omega_0}{\Omega_1}+e^{-i\omega_nt}\right]\exp\left[i\eta\left(e^{-i\hat\nu_nt}\hat A+\hat A^\dag
    e^{i\hat\nu_nt}\right)\right]+H.c.\;,
\end{equation}
where $\hat\nu_n=\frac{\omega}{2}[(\hat n+2)f^2(\hat n+2)-\hat
nf^2(\hat n)]$. In this relation the function $f(\hat n)$ is given
by Eq. (\ref{ff}).\\ \indent By using the vibrational
rotating-wave approximation \cite{vogel1} and applying the
disentangling approach in \cite{garcia} for the exponential term
which appeared in equation (\ref{hint1}), the interaction
Hamiltonian (\ref{hint1}) may be written as
\begin{equation}\label{}
    \hat H_I=\hbar\Omega_1\hat S^+\left[\frac{\Omega_0}{\Omega_1}F_0(\hat n,\eta)+g(\eta)F_1(\hat n,\eta)\hat
    a\right]+H.c.\;,
\end{equation}
where the function $F_j(\hat n,\eta)\;(j=0,1)$ is defined by
\begin{equation}\label{maindef}
    F_j(\hat
    n,\eta)=\sum_{l=0}^n\frac{[g(\eta)]^{2l}}{l!(l+j)!}\frac{f(\hat n)!f(\hat n+j)!}{[f(\hat n-l)!]^2}\frac{\hat n!}{(
    \hat n-l)!}M(\hat n-l).
\end{equation}
In this equation different functions are appeared which are
defined as follows
\begin{eqnarray}
    g(\eta)&=&\frac{i}{\sqrt{\gamma}}\tan(\sqrt{\gamma}\eta),\hspace{1cm}X_n=\beta-\gamma(2n+1),\nonumber\\
    M(n)&=&e^{-\frac{X_n}{\gamma}\ln(\cos(\sqrt{\gamma}\eta))},
\end{eqnarray}
where $\gamma=\frac{1}{N}$, $\beta=\sqrt{1+\frac{1}{N^2}}$ and
$\hat n$ is an operator whose eigenvalues, $n$, refer to the
excitation energy level number inside the trap. It is worth to
note that in the limiting case $N\rightarrow\infty$ which is
equivalent to $f(n)\rightarrow 1$, the system will reduce to the
confined ion in the harmonic oscillator-shaped trap, which has
been considered in \cite{vogel1}. The function $F_j(\hat n,\eta)$,
given in Eq.
(\ref{maindef}), will reduce to its counterpart in the harmonic oscillator-shaped trap \cite{vogel1}.\\
\indent The time evolution of the system is characterized by the
master equation
\begin{equation}\label{master}
    \frac{d\hat\rho}{dt}=-\frac{i}{\hbar}[\hat H_I,\hat\rho]+\frac{\Gamma}{2}(2\hat S^-\hat\rho^{'}\hat S^+-\hat S^+\hat S^-\hat\rho-\hat\rho\hat S^+\hat
    S^-),
\end{equation}
where $\Gamma$ is the spontaneous emission rate. To account for
the recoil of spontaneously emitted photons the first term of the
damping part of the master equation contains
\begin{equation}\label{}
    \hat\rho^{'}=\frac{1}{2}\int_{-1}^1dzY(z)e^{ik_l\hat xz}\hat\rho
    e^{-ik_l\hat xz},
\end{equation}
$Y(z)$ is the angular distribution of the spontaneous emission and
$\hat\rho$ is the vibronic density operator.\\ \indent In the
long-time limit, the ion will be populated in the ground state
$|g\rangle$ as a consequence of atomic spontaneous emission. In
this case, the steady-state solution of the master equation
(\ref{master}) can be assumed to be
$\hat\rho_{ss}=|g\rangle|\psi\rangle\langle\psi|\langle g|$, where
$|\psi\rangle$ stands for the vibronic motion of the ion. The
stationary solution of Eq. (\ref{master}) can be found by setting
$\frac{d\hat\rho}{dt}=0$ and since
\begin{equation}\label{}
    \hat S^-|g\rangle\langle g|=\hat S^+\hat S^-|g\rangle\langle
    g|=|g\rangle\langle g|\hat S^+\hat S^-=0,
\end{equation}
we obtain
\begin{equation}\label{}
    [\hat H_I,\hat\rho_{ss}]=0.
\end{equation}
From this equation, we find that the vibronic state $|\psi\rangle$
satisfies the following equation
\begin{equation}\label{NLCS}
    \hat ah(\hat
    n)|\psi\rangle=\chi|\psi\rangle,\hspace{1.5cm}\chi=-\frac{\Omega_0}{g(\eta)\Omega_1}.
\end{equation}
In this equation $h(\hat n)=F_1(\hat n-1,\eta)/F_0(\hat
n-1,\eta)$.
%---------------------------------------------------------------------------------------------
\section{Nonlinear coherent states of ionic vibrational motion and their physical
properties}\label{sec3} \indent Similar to the definition of the
canonical coherent states \cite{glauber}, the coherent state of a
generalized $f$-deformed oscillator is defined as a right-hand
eigenstate of the generalized annihilation operator $(\hat A=\hat
af(\hat n))$ as follows
\begin{equation}\label{}
    \hat A|\alpha,f\rangle=\alpha|\alpha,f\rangle.
\end{equation}
Due to the appearance of nonlinear deformation function, $f(\hat
n)$, in definition of these states, they are called nonlinear
coherent states. According to this definition, vibronic state of
the ion in the steady state, Eq. (\ref{NLCS}), is a nonlinear
coherent state with

\begin{eqnarray}
f(\hat n)&=&h(\hat n)=\frac{F_1(\hat n-1,\eta)}{F_0(\hat n-1,\eta)},\nonumber\\
\alpha&=&\chi=-\frac{\Omega_0}{g(\eta)\Omega_1}.
\end{eqnarray}
Nonlinear coherent states can be expanded in terms of the usual
Fock states $(\hat n|n\rangle=n|n\rangle)$ as follows
\begin{equation}\label{}
    |\alpha,f\rangle=N_f\sum_n\frac{\alpha^n}{\sqrt{n!}f(n)!}|n\rangle,\hspace{1cm}
    N_f=\left[\sum_n\frac{|\alpha|^{2n}}{n![f(n)!]^2}\right]^{-\frac{1}{2}},
\end{equation}
where $f(n)!=f(n)f(n-1)\cdots f(0)$. Thus, the steady state of the
ion in Eq. (\ref{NLCS}) is a special kind of the nonlinear
coherent state where its properties are defined by the function
$h(\hat n)$. This function is characterized by the Lamb-Dicke
parameter $\eta$ and quantum number $n$ which refers to the level
of vibronic excitation. Moreover, according to the Eq.
(\ref{NLCS}), nonlinear coherent state of the ion depends on the
complex parameter $\chi$, which is controlled by the Rabi
frequencies of the lasers, the Lamb-Dicke parameter and $\gamma$
parameter that governed by the range of the trap. In order to get
some insight about physical properties of this family of nonlinear
coherent state, we consider some statistical properties of this
state. In Fig. (\ref{f1}) we show the vibrational number
distribution of this state, $p(n)=|\langle n|\psi\rangle|^2$. In
all of the plots in this figure, Lamb-Dicke parameter and the
ratio $\frac{\Omega_0}{\Omega_1}$ are chosen as $\eta=0.22$ and
$\frac{\Omega_0}{\Omega_1}=0.85$, respectively. It can be seen
that the vibrational number distribution depends sensitively on
the depth (or range) of the trap. In some cases it is possible to
prepare a superposition of several Fock states. Another feature of
this figure is that by choosing the proper values of the depth of
the trap, such as $(N=30)$, it is possible to prepare a
superposition of two or three Fock states. An interesting property
of this vibrational number distribution is that we can prepare a
highly excited Fock state for external motion of the ion $(N=45)$
\cite{vogel2}. In this case with most probability we can claim
that one Fock state is prepared. By increasing the depth of the
trap $(N=75)$, the vibrational number distribution will reduce to
a superposition of Fock states again. In this case the
distribution of the Fock states is approximately symmetric about
the most probable number state. Thus, it is shown that for
definite values of physical parameters, $\eta$ and
$\frac{\Omega_0}{\Omega_1}$ and for different values of the trap
depth, we can prepare different states even a highly excited Fock
state.\\ \indent
 In Fig. (\ref{f2}) we have plotted quadrature
squeezing of the state $|\psi\rangle$, Eq. (\ref{NLCS}). Physical
parameters for this plot are chosen as
$\eta=0.25,\;\frac{\Omega_0}{\Omega_1}=0.31$ and the phase of the
quadrature operator is chosen as $\frac{\pi}{4}$. This figure
depicts squeezing behavior versus the depth of the trap. It is
evident that for some values of the depth, the state (\ref{NLCS})
exhibits quadrature squeezing. Hence, in addition to the
remarkable properties of the vibrational number distribution, this
state has other nonclassical property. The non-classical
properties of nonlinear coherent states is one of their most
important properties \cite{mancini}.\\ \indent In Fig. (\ref{f3}),
we have shown the contour plots of the $Q$ function of the state
(\ref{NLCS}). In this figure, different plots belong to different
depths of the trap with $\eta=0.75$ and
$\frac{\Omega_0}{\Omega_1}=0.9$. In the case of $N=7$ (plot (a))
the plot contains contribution at several amplitudes. This feature
implies occurrence of quantum interference effects inherent in
this state. It displays several localized regions where it becomes
extremely small. This phenomena is related to the separate peaks
of the number distribution of state (\ref{NLCS}) which are rather
close together. By increasing the depth of the trap, in plot (b),
$N=26$, and plot (c), $N=45$, this strong structure of the $Q$
function is disappeared. In these cases the $Q$ function has one
peak and this shows that the peaks of the number distribution are
decreased. On the other hand, the cross section of the $Q$
function is not symmetric and this shows that for selected values
of the parameters, the associated quadrature operator exhibits
quadrature squeezing. With more increasing the depth of the trap,
in plot (d), $N=75$, structure of the $Q$ function becomes
stronger than plots (b) and (c). In this case, the state exhibits
quadrature squeezing and we expect that quantum interference
occurs again.\\ \indent
 To obtain more information
about the nature of the state (\ref{NLCS}), we have considered its
associated Wigner function, $W(\alpha)$. The Wigner function for
different values of the Lamb-Dicke parameter and the depth of the
trap is shown in Fig. (\ref{f4}). In this figure the ratio
$\frac{\Omega_0}{\Omega_1}$ is chosen equal to $0.9$. The negative
values of the Wigner function are a signature of the nonclassical
nature of the associated state. As is seen, in all cases the
Wigner function has negative values. To consider the Lamb-Dicke
parameter effects, in plots \ref{f4}(a)-\ref{f4}(c), we have
decreased the Lamb-Dicke parameter while the depth of the trap is
chosen constant. The Wigner function in plot \ref{f4}(a) shows
occurrence of the quantum interference. Decreasing of the
Lamb-Dicke parameter splits peaks of the Wigner function in two
groups. This yields a coherent superposition of two quantum
states. It is evident that decreasing of the Lamb-Dicke parameter
will decrease the amplitude of the Wigner function. In addition to
the Lamb-Dicke parameter effects, dependence of the Wigner
function to the depth of the trap is considered in plots
\ref{f4}(d)-\ref{f4}(f). It is seen that in plot \ref{f4}(d), for
selected parameters, the Wigner function is split into two parts
which is signature of superposition of two coherent states,
because each part consists of several peaks. By increasing the
depth of the trap, these two parts are going to be mixed and the
quantum interference will be occurred.

%---------------------------------------------------------------------------------------------
\section{Conclusion}\label{sec4}
\indent We have studied dynamics of a single trapped ion in a
finite  range trap. In the context of the $f$-deformed
oscillators, we have shown that the confined ion in a finite range
trap can be assumed as an $f$-deformed oscillator. By modelling
the realistic potential with the modified P\"{o}schl-Teller
potential, we have obtained dynamical variables (raising and
lowering operators) of this system. Moreover, we have proposed a
scheme for preparation of a special family of nonlinear coherent
states. Such states could be generated as stationary states of the
center-of-mass motion of a laser-driven trapped ion in a finite
range trap while interacts with a bichromatic laser field. When
the motional state is nonlinear coherent state, the ion is
decoupled from the driving laser field. Then, any perturbation of
this motional state leads to the switching of the interaction and
this leads to a self-stabilization of the state. We have shown
that the prepared motional state of the ion has some nonclassical
features which strongly depend on the depth of the trap. These
states show some coherence effects such as localization of their
phase-space distribution and splitting to two or more sub-states
which the latter leads to quantum interference. According to the
profile of the $Q$ function of these states, they exhibit
quadrature squeezing and for specific values of the physical
parameters we have calculated their quadrature squeezing. It is
shown that the nonclassical nature of the prepared states depends
on the depth of the trap so that for specific values of the depth,
both quantum interference and quadrature squeezing will occur but
for some other values, this state exhibits quadrature squeezing only.\\
\indent In view of interesting properties of generated states in
this paper, states of this type and physical system under
consideration might to be of more general interest. First of all,
the single trapped ion in finite range trap has a finite
dimensional Hilbert space. As mentioned before, the number of
energy levels in this system is controlled by the depth of the
trap. As we know, size of the Hilbert space (dimension of the
Hilbert space) has a crucial importance in some quantum phenomena,
such as decoherence. Due to the development in experimental set
ups of trapped ion, it seems possible to organize an experiment to
consider Hilbert space size effects for this system. Then, our
system can be considered as an experimental set up to investigate
Hilbert space size effects. Second, this system turn out to be of
interest for realization of the quantum groups. If we take a look
at Hamiltonian (\ref{hint}), it seems that in the Lamb-Dicke
regime $(\eta\ll 1)$, this system can be considered as a
realization of a deformed Jaynes-Cummings model. By considering
the Lamb-Dicke limit, the exponential in Eq. (\ref{hint}) can be
expanded to lowest order, resulting in the operator $g'(\hat A\hat
S^++\hat A^{\dag}\hat S^-)$, which corresponds to the deformed
Jaynes-Cummings model (in this relation $g'=\eta g$). In addition,
it is shown that there is a relation between the operators $\hat
A$ and $\hat A^\dag$ in Eq. (\ref{a mpt}) and the $q$-deformed
algebra \cite{Davoudi}. Therefore, our model can be considered as
a realization of $q$-deformed and general deformed Jaynes-Cummings
model where Lamb-Dicke parameter plays an important role on this
issue. Third, in recent types of the Paul traps, the so-called
nano Paul traps \cite{nanotrap}, the finite range effects of
trapping potential are more important. It seems that our model
which tries to consider finite range effects can provide a
theoretical description for investigating the nano Paul traps. To
put every things in a nut shell, our model in this paper provides
an experimental set up to consider Hilbert space size effects and
realization of $q$-deformed and general $f$-deformed algebras.
%--------------------------------------------------------------------------------------
\\{\bf Acknowledgments}\\
The authors wish to thank The Office of Graduate Studies and
Research Vice President of The
    University of Isfahan for their support.

%========================================================================================

\newpage

{\bf Figure captions}\\

Fig. 1. The vibrational number distribution is shown for four
values of the depth of the trap. The values of the depth, $N$, are
written on each plot and $\eta=0.22$ and
 $\frac{\Omega_0}{\Omega_1}=0.85$.\vspace{1cm}

Fig. 2. Plot of quadrature squeezing versus depth of the trap. In
this plot $\eta=0.25$, $\frac{\Omega_0}{\Omega_1}=0.31$ and
quadrature operator phase is selected as $\frac{\pi}{4}$.
\vspace{1cm}

Fig. 3. Contour plots of the $Q$ function for $\eta=0.75$ and
$\frac{\Omega_0}{\Omega_1}=0.9$.
 In this figure light region indicates large values of the function. Each plot belongs to
 specific values of the depth of the trap. In plot (a) $N=7$, plot (b) $N=26$, plot (c) $N=45$
 and in plot (d) the depth of the trap is selected as
 $N=75$. \vspace{1cm}

Fig. 4. Plots of the Wigner function for different values of the
 Lamb-Dicke parameter and the depth of the trap which are shown on
 each plot. In all plots the ratio $\frac{\Omega_0}{\Omega_1}$ is selected equal to $0.9$.

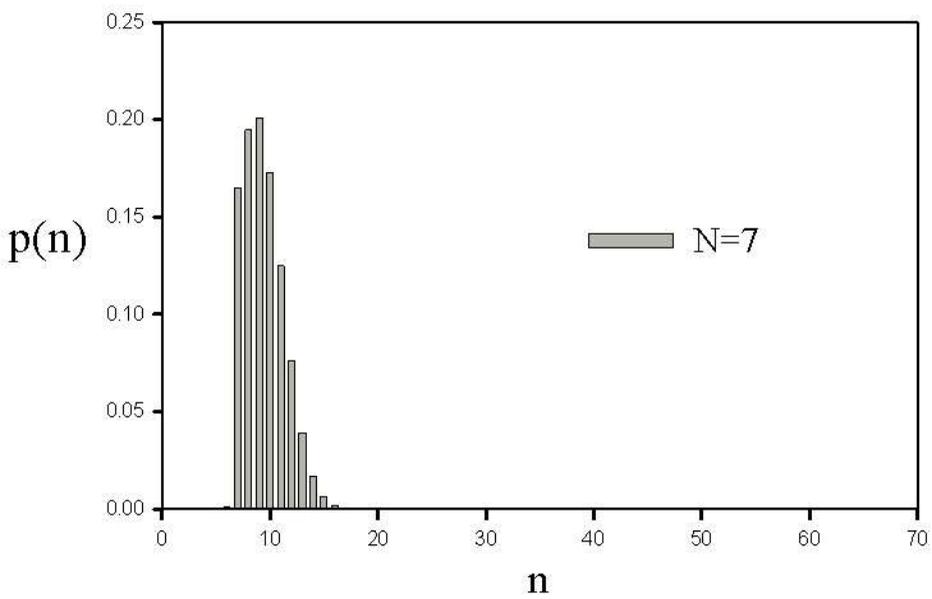
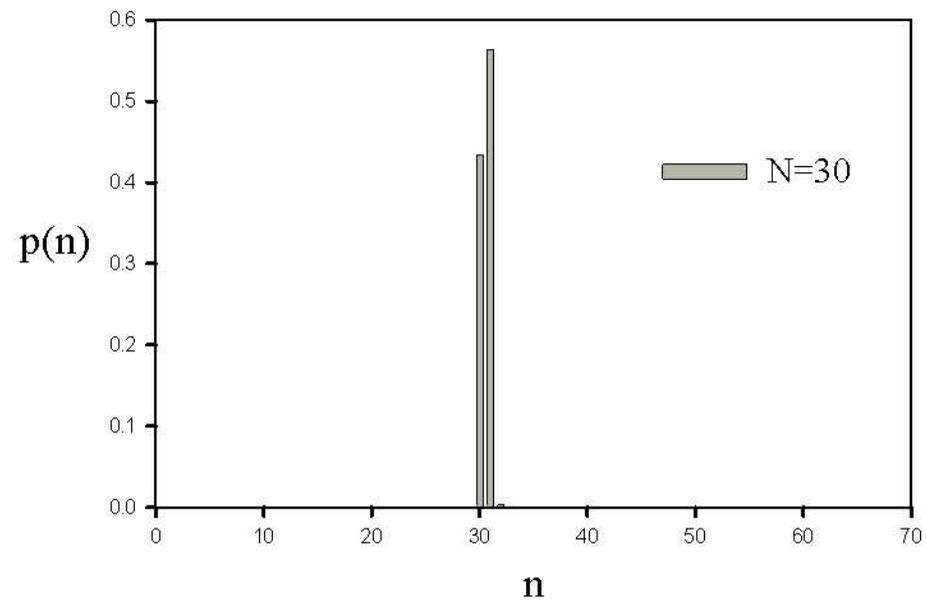
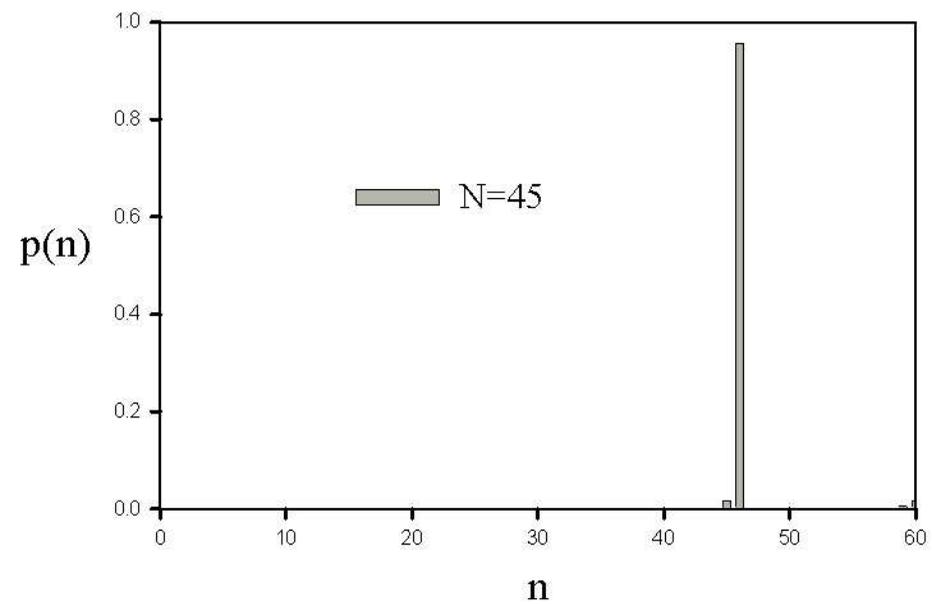
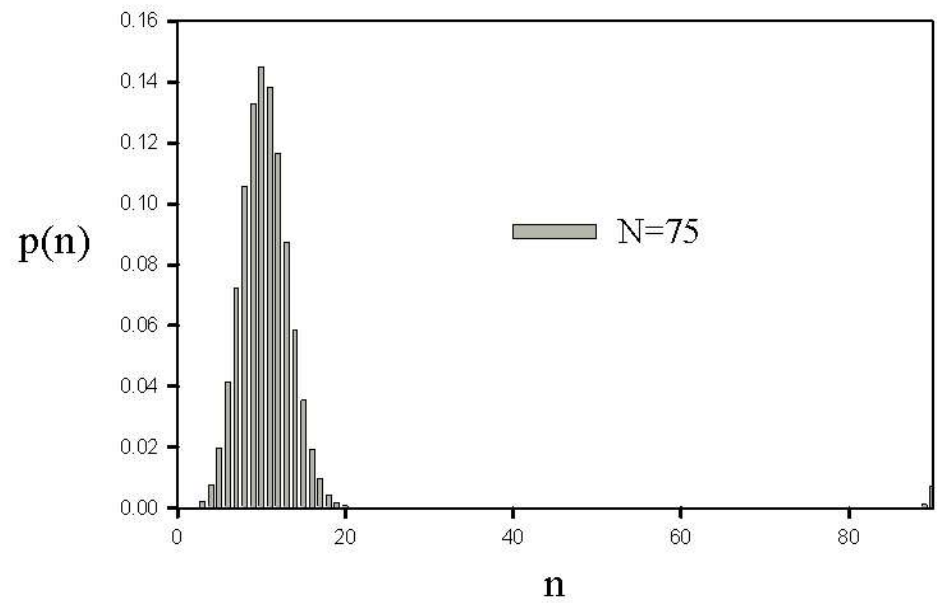
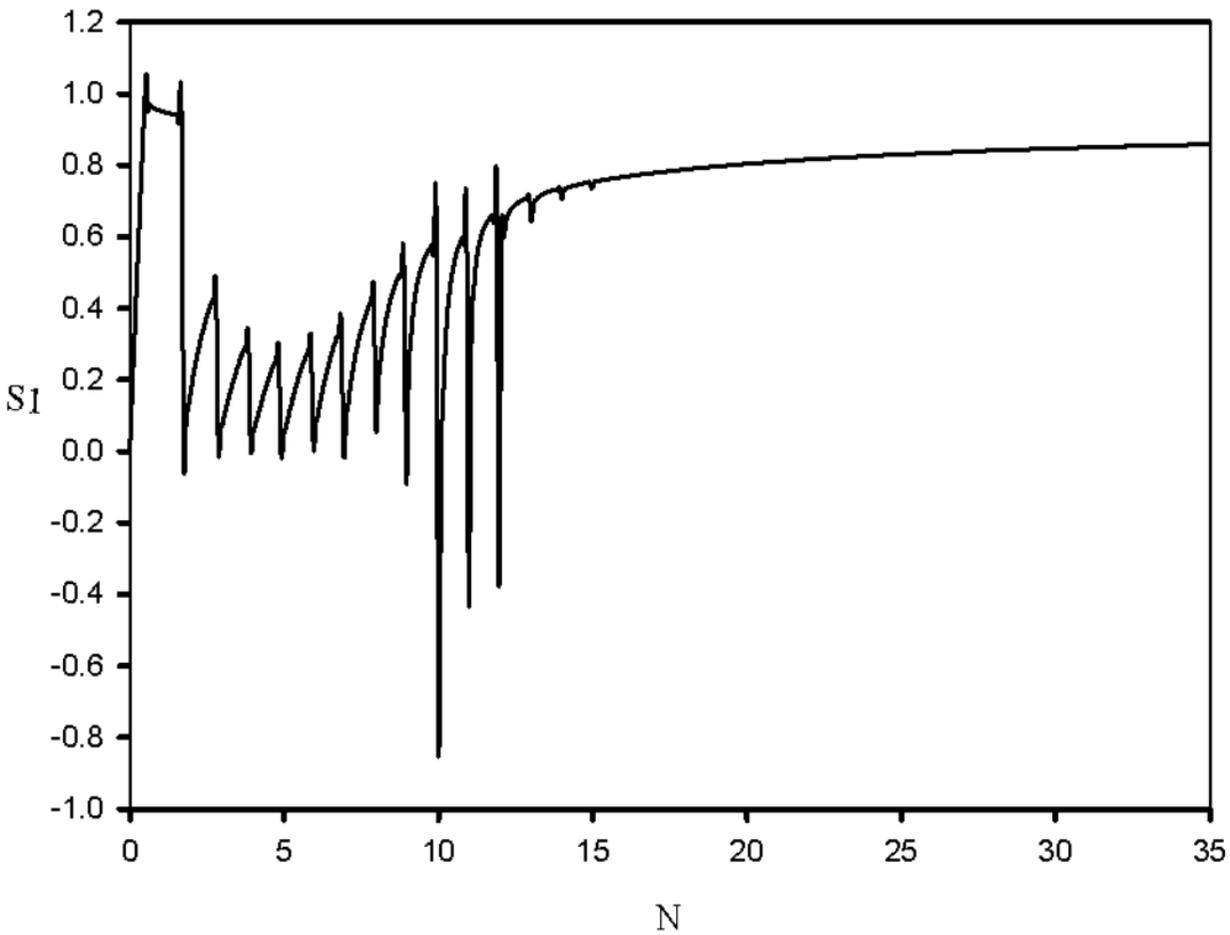
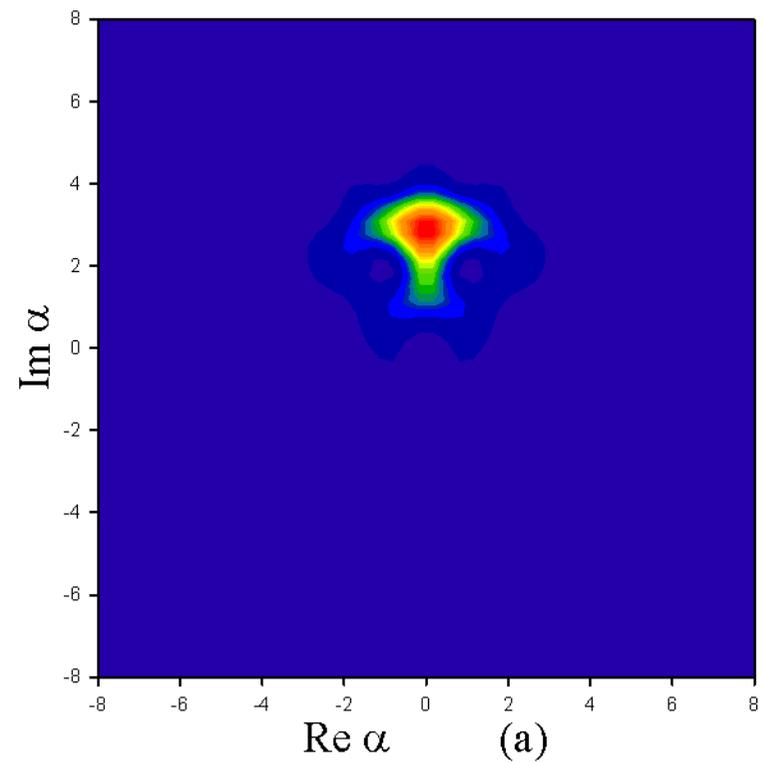
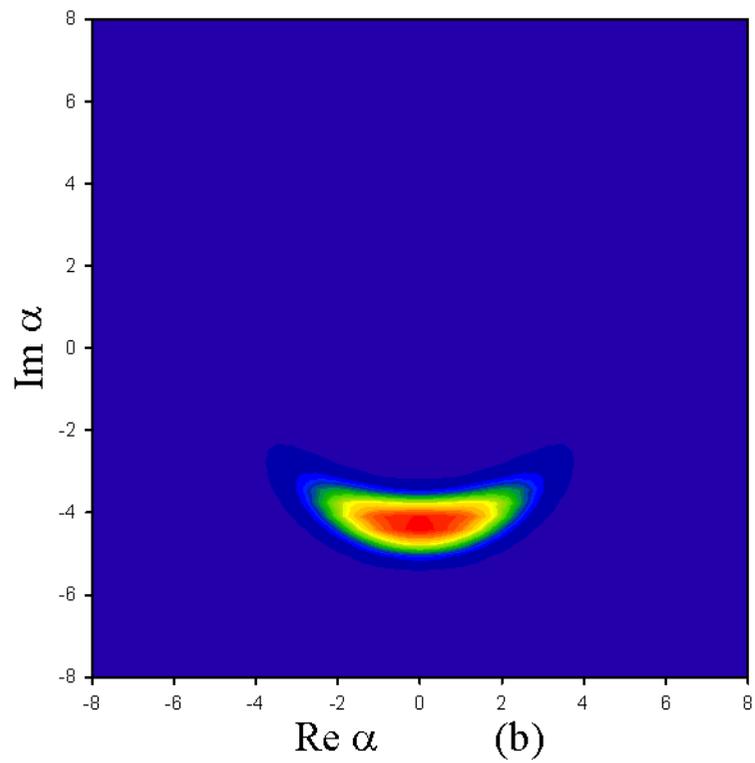
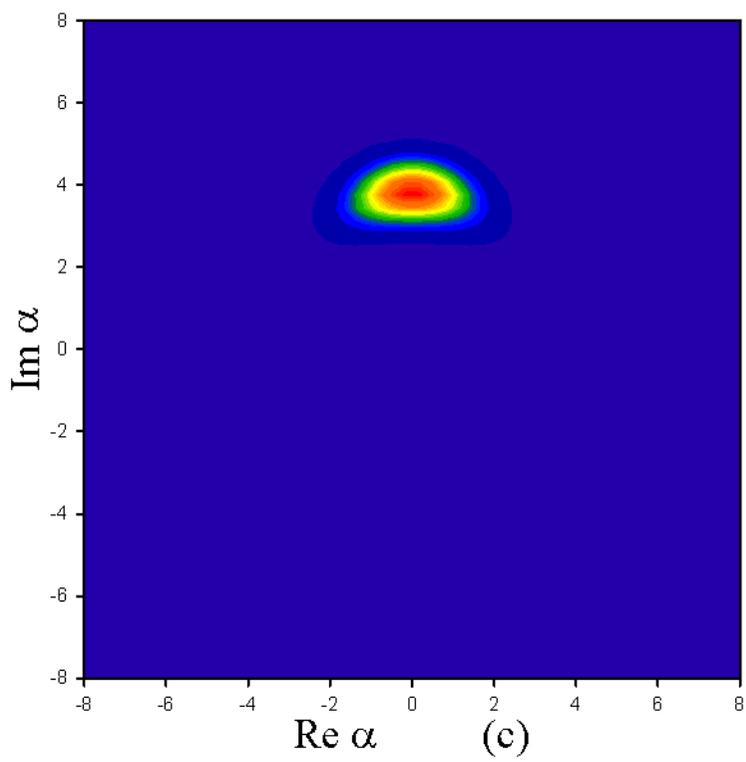
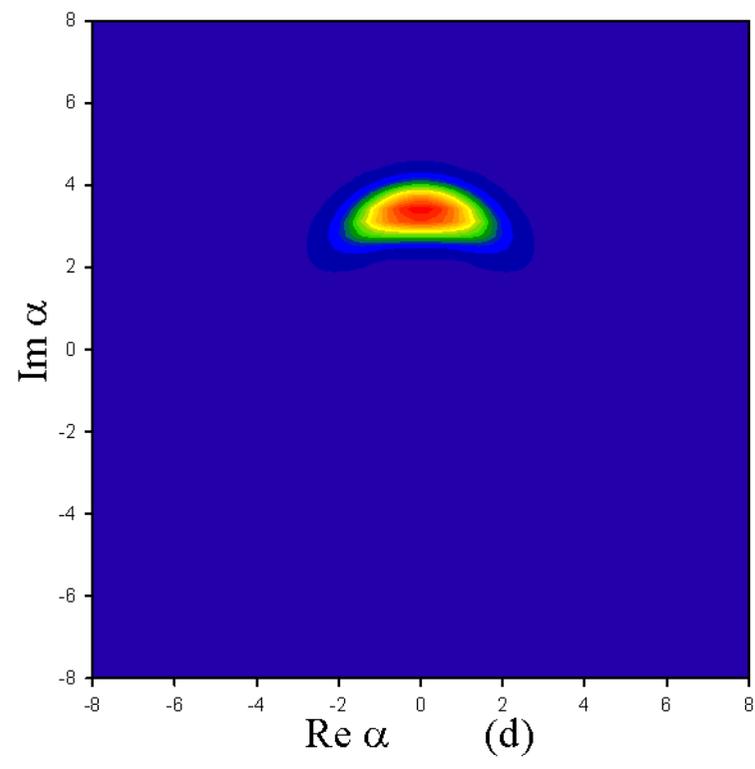
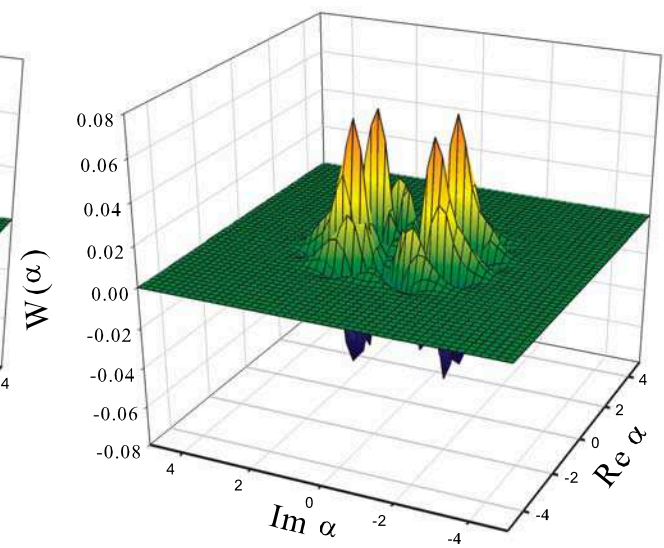
\end{document}